\documentclass{amsart}

\newcommand{\beqa}{\begin{eqnarray*}}
\newcommand{\eeqa}{\end{eqnarray*}}
\newcommand{\beqn}{\begin{eqnarray}}
\newcommand{\eeqn}{\end{eqnarray}}

\newcommand{\e}{\varepsilon}

\newcounter{cnt1}
\newcounter{cnt2}
\newcounter{cnt3}
\newcommand{\blr}{\begin{list}{$($\roman{cnt1}$)$}
 {\usecounter{cnt1} \setlength{\topsep}{0pt}
 \setlength{\itemsep}{0pt}}}
\newcommand{\bla}{\begin{list}{$($\alph{cnt2}$)$}
 {\usecounter{cnt2} \setlength{\topsep}{0pt}
 \setlength{\itemsep}{0pt}}}
\newcommand{\bln}{\begin{list}{$($\arabic{cnt3}$)$}
 {\usecounter{cnt3} \setlength{\topsep}{0pt}
 \setlength{\itemsep}{0pt}}}
\newcommand{\el}{\end{list}}
\newtheorem{thm}{Theorem}
\newtheorem{lem}[thm]{Lemma}

\newtheorem{Def}[thm]{Definition}

\newtheorem{rem}[thm]{Remark}
\newcommand{\Rem}{\begin{rem} \rm}
\newcommand{\bdfn}{\begin{Def} \rm}
\newcommand{\edfn}{\end{Def}}

\newcommand{\ba}{\begin{array}}
\newcommand{\ea}{\end{array}}

\date{}
\begin{document}
\title{\bf Adjoint for Operators in Banach Spaces}
\author[Gill]{T. L. Gill}
\address[Tepper L. Gill]{ Department of Electrical Engineering Howard University\\
Washington DC 20059 \\ USA, {\it E-mail~:} {\tt tgill@howard.edu}}
\author[Basu]{S. Basu}
\address[Sudeshna Basu]{ Department of Mathematics\\ Howard University\\
Washington DC 20059 \\ USA, {\it E-mail~:} {\tt sbasu@howard.edu}}
 
\author[Zachary]{W. W. Zachary}
\address[Woodford W. Zachary]{ Department of Electrical Engineering \\ Howard
University\\ Washington DC 20059 \\ USA, {\it E-mail~:} {\tt
wwzachary@earthlink.net}}
\author[Steadman]{V. Steadman}
\address[V. Steadman]{Department of Mathematics\\
 University of Distrcit of Columbia\\ Washington DC}
\date{}
\subjclass{Primary (45) Secondary(46) }
\keywords{Adjoints, Banach space embeddings, Hilbert spaces}
\maketitle
\begin{abstract}  In this paper we show that a result of Gross and Kuelbs, used to study Gaussian measures on Banach spaces, makes it possible to construct an adjoint for operators on separable Banach spaces. This result is used to extend well-known theorems of von Neumann and Lax. We also partially solve an open problem on
the existence of a Markushevich basis with unit norm and prove that all closed densely defined linear operators on a separable Banach space can be
approximated by bounded operators.  This last result extends a theorem of
Kaufman for Hilbert spaces and allows us to define a new metric for closed
densely defined linear operators on Banach spaces.  As an application, we
obtain a generalization of the Yosida approximator for semigroups of operators.
\end{abstract}
\section*{Introduction} One of the greatest impediments to the development of a
theory of operators on Banach spaces that parallels the corresponding theory on
Hilbert spaces is the lack of a suitable notion  of an adjoint operator for
these spaces. It is an interesting fact of history that the tools needed were
being developed in probability theory during the time of greatest need.
It was in 1965, when Gross \cite{G} first proved that every real separable
Banach space contains a separable Hilbert space as a dense embedding, and this
(Banach) space is the support of a Gaussian measure.
 Gross'  theorem was a far reaching generalization of Wiener's theory, which
was based on the use of the (densely embedded Hilbert)  Sobolev space
${\bf H^{1}}[0,1]\subseteq {\bf C}[0,1].$ Later, Kuelbs
\cite{K} generalized Gross' theorem to include the fact that
${\bf H^{1}}[0,1]\subseteq {\bf C}[0,1]\subseteq {\bf L^{2}}[0,1].$ This
Gross-Kuelbs theorem can be stated for our purposes as:
\begin{thm} {\rm (Gross-Kuelbs)} Suppose $\bf{B}$ is a separable Banach space.  Then there exist separable Hilbert spaces $\bf {H_1, H_2}$ and a positive trace class operator  $\bf{T_{12}}$ defined on $\bf{H_2}$ such that $\bf{H_1}\subseteq \bf{B} \subseteq \bf{H_2}$ (all as continuous dense embeddings), and $\bf{T_{12}}$ determines $\bf{H_1}$ when $\bf{B}$ and $\bf{H_2}$ are given.
\end{thm}
\section*{Purpose} The purpose of this paper is to show that the Gross-Kuelbs
theorem makes it possible to give an essentially unique definition of the
adjoint for operators on separable Banach spaces. This definition has all the
expected properties. In particular, we show that, for each bounded linear
operator $\bf{A}$, there exists $\bf{A^*}$, with $\bf{A^{*}A}$ maximal
accretive, self adjoint
$\bf{(A^{*}A)^*  = A^{*}A}$,   and $\bf{I+A^{*}A}$ is invertible.
Although our main interest is in the construction of a generalized Yosida
approximator for semigroups of operators that will be used elswhere, this
adjoint has a number of important implications for other aspects of operator
theory.  As a sampling, we provide generalizations of theorems due to von
Neumann \cite{VN}, Lax \cite{L}, and Kaufman \cite{Ka} to Banach spaces. We
also partially solve an open problem on the existence of a Markushevich basis
with unit norm.
\section*{Background} In what follows, we let $L[\bf{B}]$, $L[\bf{H}]$ denote
the bounded linear operators on $\bf{B}$, $\bf{H}$ respectively. By a duality
map, $\phi_x$, defined on $\bf{B}$, we mean any linear functional
$\phi_{x} \in \{f \in {\bf B^{'}}| <x,f> = \|x\|^{2}_{\bf B}, x \in {\bf B}\}$,
where $<.>$ is the natural pairing between a Banach space and its dual. Let
$\bf{J}: \bf{H} \longrightarrow \bf{H'}$ be the standard conjugate isomorphism
between a
Hilbert space and its dual, so that ${<x, {\bf J}(x)> = (x,x)_{\bf H}=
\|x\|^{2}_{\bf H}}.$ We define the {\it special duality map} of $\bf{B}$
associated with $\bf{H}$ by:
 
$$\phi^{s}_{x}= \frac {\|x\|^{2}_{B}}{\|x\|^{2}_{\bf H}} {\bf J}(x). $$
It is easy to check that $\phi_{x}^{s}$ is a duality map for $\bf{B}$. A closed
densely defined operator  $\bf{A} $ is called maximal accretive if $ <{\bf
A}x, \phi_{x}> \geq 0$ for all $x \in D(\bf{A})$ and $\bf{A}$ has no proper
extension. The following results due to von Neumann \cite{VN} and  Lax \cite{L}
are listed for reference.
 
\begin{thm}{\rm (von Neumann)} For any closed densely defined linear operator
$\bf{A}$ on a Hilbert space $\bf{H}$, the operators $\bf{A^{*}A}$ and $\bf{I +
A^{*}A}$ are selfadjoint, and $\bf{I+ A^{*}A}$ has  a bounded inverse.
\end{thm}
\begin{thm}{\rm (Lax)} Let $\bf{H_2}$ be given so that $\bf{B} \subseteq
\bf{H_2}$ densely.  If $\bf{A}$ is a bounded linear operator on $\bf{B}$ such
that $\bf{A}$ is  selfadjoint (i.e., $({\bf A}x,y)_{\bf{H_2}}~ = ~(x ,{\bf
A}y)_{\bf H_{2}}\quad \forall x, y, \in \bf{B}$ ), then $\bf{A}$ is bounded on
${\bf H_2}$ and $\|{\bf A}\|_{\bf{H_2}}\leq \|{\bf A}\|_{\bf{B}}$.
\end{thm}
\section*{Main Results} Let us fix $\bf{H_{1}}, \bf{H_{2}}$ such that
$\bf{H_{1}} \subseteq \bf{B} \subseteq \bf{H_{2}}$ as continuous dense
embeddings, and, without loss of generality, assume that for $x \in \bf{H_1}$,
$\|x\|_{2} \leq \|x\|_{\bf{B}} \leq \|x\|_{1}.$
 
The first result is not new and is, in fact, well known. We present it because
the proof is new and uses specific information about the relationship between
$\bf{B}$ and $\bf{H_2}$.
\begin{thm} Every closed densely defined linear operator on
${\bf B}$ extends to a closed densely defined linear operator on %
${\bf H_2}.$
\end{thm}
\begin{proof}
 Let ${\bf J_2}:{\bf H_2}\longrightarrow {\bf H'_2}$
 denote the standard conjuate isomorphism. Then, as $ {\bf B}$ is strongly
dense in $ {\bf H_2}$,
${\bf J_2}[{\bf B}] \subset {\bf H'_{2}} \subset {\bf B'}$ is (strongly) dense
in  ${\bf H'_{2}}.$ If
${\bf A}$ is any closed densely defined linear operator on ${\bf B}$ with
domain $D({\bf A}),$ then ${\bf A'}$ (the ${\bf B}$ adjoint of ${\bf A}$) is closed on ${\bf B'}.$ In addition,
${\bf A^{'}|}_ {\bf H'_{2}} $ is closed and, for each
$x \in D({\bf A})$, ${\bf J_2}(x) \in {\bf H'_{2}}$
 and
$< {\bf A}y,{\bf J_2}(x) >$ is well defined  $\forall y \in D({\bf A}).$
 Hence ${\bf J_2}(x) \in D({\bf A'})$ for all $x \in D ({\bf A})$.
 Since ${\bf J_2}({\bf B})$ is strongly dense in  ${\bf H'_{2}}$, this implies
that
${\bf J_2}(D({\bf A'})) \subset D({\bf A'})$ is strongly dense in
${\bf H'_{2}}$ so that $D({\bf A'})\left| _{{\bf H'_{2}} } \right.$ is strongly
dense in ${\bf H'_{2}}.$ Thus, as ${\bf H_{2}}$ is reflexive,
$
\left[ {{\bf A'}\left| {_{{\bf H'_{2}}} } \right.} \right]^\prime$ is a closed
densely defined operator on
${\bf H_{2}}.$
\end{proof}
 
In the next theorem, we prove that every bounded linear operator
${\bf A}$ on ${\bf B}$ has a well defined adjoint.  The result is actually true
for any closed densely defined linear  operator on ${\bf B}$ but, in this case,
for each
${\bf A}$ we must have
${\bf H}_1  \subseteq D({\bf A})$ so, in general, a different ${\bf H}_1 $
 is required for each operator.  It should also be noted that, although ${\bf H_1}$ and ${\bf H_2}$ are
required to obtain our adjoint, it is not hard to show that any two adjoint
 operators for ${\bf A}$ will differ by a similarity transformation of unitary
operators (see Theorem $11$).
\begin{thm}  Let
${\bf B}$ be a separable Banach space with
${\bf A} \in L[{\bf B}].$ Then there exists
${\bf A}^ * \in L[{\bf B}]$ such that:
\begin{enumerate}
\item
${\bf A}^ *  {\bf A}$ is maximal accretive.
\item
$({\bf A}^ *  {\bf A})^ *   = {\bf A}^ *  {\bf A}$, and
\item
${\bf I} + {\bf A}^ *  {\bf A}$ has a bounded inverse.
\end {enumerate}
\end{thm}
\begin{proof}  If we let ${\bf J}_i :{\bf H}_i  \to {\bf H'}_i $, $(i =1, 2)$,
then
${\bf A}_1 \quad   =  \quad {\bf A}_{|{\bf H}_{1}  }: {\bf H}_1
\longrightarrow {\bf H}_2 ,$ and ${{\bf A'}_1 :{\bf H'}_2  \longrightarrow {\bf
H'}_1}.$
 
It follows  that ${\bf A'}_1 {\bf J}_2 :{\bf H_2}  \longrightarrow {\bf H'}_1 $
and ${\bf J}_1^{ - 1} {\bf A'}_1 {\bf J}_2 :{\bf H}_2  \to {\bf H}_1  \subset
{\bf B}$ so that, if we define
${\bf A}^ {*}   = [ {{\bf J}_1^{ - 1} {\bf A'}_1 {\bf J}_2 } ]_{\bf B},
$ then ${\bf A}^ *  :{\bf B} \to {\bf B}$ (i.e., ${\bf A}^ *   \in L[{\bf
B}]$).
To prove 1, ${\bf J'}_i  = {\bf J}_i $ and, if $x \in {\bf B}$, then $
\langle {{\bf A}^ *  {\bf A}x,{\bf J}_2 (x)}\rangle  = \langle {{\bf A}x,({\bf
A}^ *   )'{\bf J}_2 (x)} \rangle .$ Using the above definition of
${\bf A}^ * $, we get that
$({\bf A}^ * ){'}{\bf J}_2 (x) = \{ [ {\bf J}_1^{ - 1} {\bf A'}_1 {\bf J}_2 ]_
{ \bf B } \}^{'}  {\bf J}_2 (x) = [ {\bf J}_2 {\bf A}_1 {\bf J}_1^{ - 1}  ]{\bf
J}_2 (x) = {\bf J}_2 ({\bf A}_1 x).$ Since, for $x \in {\bf H}_1$,
${\bf A}_{1} x = {\bf A}x$ and
$$
\langle {{\bf A}^ * {\bf A}x, \phi _x^s } \rangle  = \frac{\| x \|_{\bf B}^2 }
{\| x \|_2^{2} }\langle {{\bf A}x,{\bf J}_2 ({\bf A}_1 x)} \rangle  =
\frac{{\| x \|_{\bf B}^2 }} {{\| x \|_2^2 }}\| {{\bf A}x} \|_2^2  \geq 0,$$
we
have that
${\bf A}^ *  {\bf A}$ is accretive on a dense set.   Thus,
${\bf A}^ *  {\bf A}$ is accretive on ${\bf B}.$ It is maximal accretive
because it has no proper extension.
To prove 2,   we have that  for
$x \in {\bf H}_1,$
\beqa ({\bf A}^ *  {\bf A})^ *  x&  = & ( {\{ {{\bf J}_1^{ - 1} [ {\{ {[ {{\bf
J}_1^{  - 1} {\bf A'}_1 {\bf J}_2 } ]| {_{\bf B} }{\bf A}} \}_1 } ]^\prime
{\bf  J}_2 } \}| {_{\bf B} } } )x  \\
  &  = &( {\{ {{\bf J}_1^{ - 1} [ {\{ {{\bf A'}_1 [ {{\bf J}_2 {\bf  A}_1 {\bf
J}_1^{ - 1} } ]| {_{\bf B} } } \}} ]{\bf J}_2 } \}|  {_{\bf B} } } )x \\ &=
&{\bf A}^ *  {\bf A}x.
\eeqa
It follows that the same result holds on ${\bf B}.$
Finally, the proof that  ${\bf I} + {\bf A}^ *  {\bf A}$
 is invertible follows the same lines as in von Neumann's theorem.
\end{proof}
\begin{thm} Every bounded linear operator on
${\bf B}$
 extends to a bounded linear operator on ${\bf H}_2$ and
$\| {\bf A} \|_{{\bf H}_2 }^2  \leq{C}\| {\bf A} \|_{\bf B}^2 $  for some
constant C.
\end{thm}
\begin{proof}: For any bounded linear operator
${\bf A}$ defined on ${\bf B}$, let
${\bf T} = {\bf A}^ *  {\bf A}$.  By Theorem 1, ${\bf T}$ extends to a closed
linear operator ${\bf T}$ on ${\bf H}_2.$  As ${\bf T}$ is selfadjoint on
${\bf B}$ , by Lax's theorem, ${\bf T}$ is bounded on ${\bf H}_2 $ and $ \|{\bf
A}^ * {\bf A}\|_{{\bf H}_2 } = \| {\bf A} \|_{{\bf H}_2  }^2 \, \leq \| {{\bf
A}^ *  {\bf A}} \|_{\bf B} \, \leq C \| {\bf A}
\|_{\bf B}^2$, where
$C = inf \{ M |~  \| {\bf A}^ *  {\bf A} \|_{\bf B}
\leq M  \| {\bf A}\|_{\bf B}^2  \}.$
\end{proof}
It should be noted that, in general,
$\| {{\bf A^* A}} \|_{\bf B}  \neq  \| {\bf A} \|_{\bf B}^2 $ and
$({\bf A} {\bf B})^*x  \neq  {\bf B^*} {\bf A^*}x .$ Thus, as
expected, there are some important differences compared to the corresponding
operator results in Hilbert spaces.  On the other hand, we can give
a  natural definition of orthogonality for subspaces of a Banach space.
\begin{Def}  Let ${\bf U}$ and  ${\bf V}$ be subspaces of ${\bf B}.$  We say
that ${\bf U}$ is orthogonal to ${\bf V}$ if,
$\forall x \in {\bf U},\langle {y,\varphi _x^s } \rangle  = 0 \quad \forall y
\in {\bf  V}.$
\end{Def}
 
The above definition is transparent if we note that
$\langle {y,\phi _x^s } \rangle  = 0 \quad  \forall y \in {\bf V}
\Leftrightarrow
\langle {y,J_2 (x)} \rangle  = 0 \quad \forall y \in {\bf V}.$
The next result is easy to prove.
\begin{lem}  If ${\bf U}$ is orthogonal to
${\bf V}$,  then ${\bf V}$ is orthogonal to
${\bf U}.$
\end{lem}
\begin{Def} A biorthogonal system $\{ {x_n , x_n^ *} | n \geq 1 \}$ is called a
Markushevich basis for  ${\bf B}$ if the span of the
$x_n $ is dense in ${\bf B}$ and the span of the
$x_n ^ {* } $ is weak* dense in  ${\bf B'}$.
\end{Def}
Pelczynski \cite{P} has shown that, for every separable Banach space
${\bf B}$ and each $\epsilon  > 0$,
${\bf B}$ has a Markushevich basis such that
$\| {x_n } \| \| {x_n^ *  } \| \leq1 + \epsilon. $ Diestel (\cite{D}, pg. 56)
notes that the question of whether  it is possible to require that  $\| {x_n }
\|\, = 1 = \| {x_n^ *  } \|$ is open.  In the next theorem, we show that, if
${\bf B}$ has a basis for a dense subspace, it has a Markushevich basis with unit norm.
\begin{thm} Let
${\bf B}$ be a separable Banach space with a basis for a dense subspace. If this basis is
normalized and monotone with respect to the ${\bf B}$ norm, then ${\bf B}$ has a
Markushevich basis $\{ {x_n ,x_n^ *  }| n \geq1 \}$ such that $
\| {x_n } \|_{\bf B}  = 1 = \| {x_n^ *  } \|_{\bf B'}.$
\end{thm}
\begin{proof} (A basis is monotone if ${y = \sum {a_i x_i }}$,  then
${\left\| {\sum\limits_{i = 1}^{m} {a_i x_i } }
\right\|_\mathbf{B}  \leq \left\| {\sum\limits_{i = 1}^{m + n} {a_i x_i } }
\right\|_\mathbf{B}}$   for  ${m, n \geq1}$.)
Let
$\{ {x_n  }| n \geq1 \}$ be a complete orthogonal basis for
${\bf H}_1 $ with  $\| {x_n } \|_{\bf B}  = 1$.
If we now define
$ x^{*}_{n} = \varphi _n^s  = \frac{{{\bf J}_2 (x_n )}} {{\| {x_n }\|_{{\bf
H}_2 }^2 }},$
 then it is easy to check that
$\langle {x_i ,x_j^ * } \rangle  = \delta _{ij}$. By definition, the span of
the family $\{ {x_n }| n \geq1\}$ is dense in ${\bf B}$
 and it is also easy to see that  the span of the family $\{x^{*}_{n}, n
\geq1\}$
 is weak* dense in  ${\bf B}^ {'}. $
To show that
$\| {x_n^ *  } \|_{\bf B}^ {'}  = 1,$
let $y=\sum\limits_{i=1}^N {a_ix_i},\ \left\| y
\right\|_B\le1,$  with $N \geq1.$ Then
${\left| {\left\langle {y,\varphi _n^s} \right\rangle } \right|\le \left| {a_n}
\right|\le \left\| y \right\|_{\bf B}\le1},$  so that
${\left\| {\varphi _n^s} \right\|_{B}=\ \mathop {sup}\limits_{\left\| y
\right\|_B\le 1}\left| {\left\langle {y,\varphi _n^s} \right\rangle }
\right|\le1}$.   We are done since
$\left\langle {x_n,\varphi _n^s} \right\rangle =1\ $.
\end{proof}
It is clear that much of the operator theory on Hilbert spaces can be extended
to separable Banach  spaces in a straightforward way.  To get a flavor, we give
a few of the more interesting results.   Since the proofs are easy, we omit
them.  In what follows, all definitions are the same as in the  case of a
Hilbert space.
\begin{thm} Let ${\bf A} \in L[{\bf B}].$
\begin{enumerate}
\item The set $N({\bf B})$ of all bounded normal operators on ${\bf B}$ is a
closed subset of
$L[{\bf B}].$
\item If ${\bf A}$ is unitary on ${\bf B},$ then there exists a selfadjoint
operator  ${\bf W}$, and  ${\bf A} = \exp (i{\bf W}).$
\end{enumerate}
\end{thm}
\section*{APPLICATION: THE YOSIDA APPROXIMATOR}
If ${\bf A}$ is the generator of a strongly continuous semigroup
$T(t) = \exp (t{\bf A})$ on ${\bf B}$, then the Yosida approximator for  ${\bf
A}$ is defined by
${\bf A}_\lambda   = \lambda {\bf A}R(\lambda ,{\bf A})$, where  $ R(\lambda
,{\bf A}) = (\lambda I - {\bf A})^{ - 1} $ is the resolvent of ${\bf A}.$ In
general,  ${\bf A}$ is closed and densely defined but unbounded.  The Yosida
approximator
${\bf A}_\lambda$ is bounded, converges strongly to ${\bf A}$, and $T_\lambda
(t) = \exp (t{\bf A}_\lambda  )$ converges strongly to $T(t) = \exp (t{\bf A}).$
If ${\bf A}$ generates a contraction semigroup, then so does
${\bf A}_\lambda $ (see Pazy \cite{Pz}).  This result is very useful for
applications.  Unfortunately, for general  semigroups, ${\bf A}$ may not have a
bounded resolvent.  Furthermore, it is very convenient to have a contractive
approximator.   As an application of the theory in the previous section, we
will show that the Yosida approach  can be generalized in such a way as to give
a contractive approximator for all strongly continuous  semigroups of operators
on
${\bf B}.$
For any closed densely defined linear operator  ${\bf A}$ on ${\bf B}$, let $
{\bf T} =  - [ {{\bf A}^ *  {\bf A}} ]^{1/2}, {\bf \bar T} =  - [ {{\bf AA}^ *
} ]^{1/2} .$ Since $
 - {\bf T}( - {\bf \bar T})$ is maximal accretive,  $ {\bf T} ({\bf \bar T})$
generates a contraction semigroup.  We can now write
${\bf A}$ as ${\bf A} = U{\bf T}$,  where $U$ is a partial isometry (since the
extension is valid on
${\bf H}_2 $, the restriction is true on  ${\bf B}$).   Define ${\bf
A}_\lambda$ by
${\bf A}_\lambda   = \lambda {\bf A}R(\lambda ,{\bf T})
$.   Note that $ {\bf A}_\lambda   = \lambda U{\bf T}R(\lambda ,{\bf T}) =
\lambda ^2 UR(\lambda ,{\bf T}) -
\lambda U$ and, although ${\bf A}$ does not commute with $ R(\lambda ,{\bf
T})$, we have $
\lambda {\bf A}R(\lambda ,{\bf T}) = \lambda R(\lambda ,{\bf \bar T}){\bf A}$.
 
\begin{thm}\label{yosida} For every closed densely defined linear operator
${\bf A}$ on ${\bf B}$, we have that
\begin{enumerate}
\item
${\bf A}_\lambda$ is a bounded linear operator and
$\mathop {\lim }_{\lambda  \to \infty } {\bf A}_\lambda  x = {\bf A}x, \forall
x \in D({\bf A}),$
\item $\exp [ {t{\bf A}_\lambda  } ]$ is a bounded contraction for $t > 0$, and
\item if ${\bf A}$ generates a strongly continuous semigroup
$T(t)={\exp [ t{\bf A} ]}$ on $D$ for $t > 0$, ${D(\bf A)} \subseteq D,$ then
$\mathop {\lim }_{\lambda  \to \infty } \|  {\exp [ {t{\bf A}_\lambda }]x} -
{\exp [ {t{\bf A}} ]x} \|_{\bf B}  = 0
\quad \forall x \in D.$
\end{enumerate}
\end{thm}
\begin{proof}:  To prove 1, let ${x \in D({\bf A})}$. Now use the fact that
$
\mathop {\lim}_{\lambda  \to \infty } \lambda R(\lambda ,{\bf \bar T})x=x$
and
$
{\bf A}_\lambda x={\lambda R(\lambda ,{\bf \bar T}){\bf A}x}.$ To prove 2,
use
$ {{\bf A}_\lambda}={\lambda ^2} UR(\lambda ,{\bf T})-{\lambda U}$,
${\| {\lambda R(\lambda ,{\bf T})} \|_{\bf B}}=1,$ and $\| U \|_{\bf B}  = 1$ to
get that
$
\| {\exp [ {t\lambda ^2 UR(\lambda ,{\bf T}) - t\lambda U} ]} \|_{\bf B}
\leq \exp [ { - t\lambda \| U  \|_{\bf B} } ]\exp [ {t\lambda {\| U
\|_{\bf B}} \| {\lambda R(\lambda ,{\bf T})} \|_{\bf B} } ] \leq1.$
 
To prove 3, let
$t > 0$ and $x \in D({\bf A}).$   Then
\beqa
 \| \exp{[ {t {\bf A}} ]x} - \exp{[ {t{\bf A}_\lambda}]x} \|_{\bf B}  & = &
\| {\int_0^t \frac{d}{ds}}[{e}^{(t - s){\bf A}_{\lambda}}{e}^{s {\bf A}}]xds
\|_{\bf B}   \\
   & \leq & \int_0^t \| [ {e}^{(t - s){\bf A}_{\lambda}} (  {\bf A} - {\bf
A}_\lambda ){e}^{s{\bf A}} x] \| _{\bf B}\\
 &  \leq &
\int_0^t {\| {[ {( {{\bf A} - {\bf A}_\lambda  } ){e}^{s{\bf A}} x} ]}
\|} _{\bf B} ds.
\eeqa
Now use
$\| {[ {{\bf A}_\lambda  {e}^{s{\bf A}} x} ]} \|_{\bf B}  =
\| [{\lambda R(\lambda ,{\bf \bar T}){e}^{s{\bf A}} {\bf A}x} ] \|_{\bf B}
\leq \| {[ {{e}^{s{\bf A}} {\bf A}x} ]} \|_{\bf B}
$ to get $
\| {[ {( {{\bf A} - {\bf A}_\lambda  } ){e}^{s{\bf A}} x} ]}
\|_{\bf B}  \leq 2\| {[ {e}^{s{\bf A}} {\bf A}x} ] \|_{\bf B}  .$
 Now, since
$
\| {[ {{e}^{s{\bf A}} {\bf A}x} ]} \|_{\bf B}
$ is continuous, by the bounded convergence theorem we have
$
\mathop {\lim }_{\lambda  {\to \infty} } \| {\exp [ {t{\bf A}} ]x - \exp
[{t{\bf A}_\lambda  } ]x} \|_{\bf B}
\leq \int_0^t {\mathop {\lim }_{\lambda  { \to \infty} } \| [ ( {\bf A} - {\bf
A}_\lambda  }  ){e}^{s{\bf A}} x ] \| _{\bf B} ds = 0.$
\end{proof}
 
\section*{CONCLUSION} The first part of Theorem \ref{yosida} is a
generalization of a result of Kaufman \cite{Ka}.  This allows us to provide a
new metric for closed densely defined linear operators on Banach spaces. If $A,
B$   are closed and densely defined,  we can define our metric by
$ d\left( {A,B} \right) = \left\| {A_0  - B_0 } \right\|,{\rm   }A_0  = A\left(
{1 + A^ *  A} \right)^{ - \frac{1}{2}} ,{\rm  }B_0  = B\left( {1 + B^ *  B}
\right)^{ - \frac{1}{2}}.
$
The Hille-Yosida Theorem for contraction semigroups gives necessary and
sufficient conditions for a closed densely defined linear operator to be a
generator.  The general strongly continuous case may be reduced to the
contraction case by shifting the spectrum and using an equivalent norm.  The
second part of Theorem 12 may be viewed as an improvement in the sense that, by
using the approximator, this procedure is no longer required.

\end{document}